# Atomistic Representation of Anomalies in the Failure Behaviour of Nanocrystalline Silicene

Tawfiqur Rakib[1], Sourav Saha[1*], Mohammad Motalab[1], Satyajit Mojumder[1], Md Mahbubul Islam[2*]

[1]Department of Mechanical Engineering, Bangladesh University of Engineering and Technology, Dhaka-1000, Bangladesh.

[2]School of Materials Engineering, Purdue University, West Lafayette, IN-47907, USA

*ssaha09@me.buet.ac.bd, *islam3@purdue.edu

**Abstract**

Silicene, a 2D analogue of graphene, has spurred a tremendous research interest in the scientific community for its unique properties essential for next-generation electronic devices. In this work, for the first time, we present a molecular dynamics (MD) investigation to determine the fracture strength and toughness of nanocrystalline silicene (nc-silicene) sheet of varied grain sizes and pre-existing crack lengths at room temperature. Our results suggest that the transition from an inverse pseudo Hall-Petch to a pseudo Hall-Petch behaviour in nc-silicene occurs at a critical grain size of 17.32 nm. This phenomenon is also prevalent in nanocrystalline graphene. However, nc-silicene with pre-existing cracks exhibits anomalous crack propagation and fracture toughness behaviour. We have observed two distinct types of failure mechanisms (crack sensitive and insensitive failure) and devised the mechano-physical conditions under which they occur. The most striking outcome, however, is that despite the presence of a pre-existing crack, the crack sensitivity of nc-silicene is found to be dependent on the grain size and their orientations. Fracture toughness calculated from both Griffith's theory and MD simulations indicate that the former over-predicts the fracture toughness of nc-silicene. This study is the first direct comparison of atomistic simulations to the continuum theories to predict the anomalous behaviour in deformation and failure mechanisms of nc-silicene.

# 1. Introduction

Silicene, a silicon-based graphene-like 2-D material[1,2], stimulated great interests among the researchers due to its outstanding mechanical and electronic properties[3–6]. Silicene is a monolayer of silicon atoms arranged in hexagonal honeycomb lattice where the atoms are not purely in $sp^2$ hybridized state but share a π bond among them[7]. Therefore, silicene has an analogous structure like graphene. However, recent studies suggested that the honeycomb lattice structure of silicene is slightly buckled due to $sp^3$ hybridization, unlike $sp^2$ hybridization in Graphene[8].

Silicene is a zero band gap semiconductor like graphene[9,10] and has tremendous applications in electronic industry[11–13]. Despite the potential of a wide range of electronic applications of silicene, its mechanical properties remain poorly understood. A great deal of research efforts has been expended to study the mechanical properties of single crystal silicene[14–16], while the mechanical properties of polycrystalline silicene remain largely unexplored. Since most of the applications involve polycrystalline structure, it makes the comprehensive study of polycrystalline silicene an evident necessity. We note that polycrystalline silicene has not yet been synthesized in laboratory. However, a recent investigation predicted the possibility of obtaining polycrystalline structure in the synthesis of silicene[17] for practical applications. That spurred a significant interest to investigate the mechanical properties of nc-silicene.

The presence of grains significantly influences mechanical properties of any materials[18,19]. The polycrystalline materials exhibit a tremendous reduction in strength compared to their single crystal due to the presence of a network of grain boundaries[20]. In polycrystalline ductile materials, the strength as a function of grain size is expressed by the following Hall-Petch relation[21,22],

$$\sigma_y = \sigma_0 + \frac{K_f}{\sqrt{d}}. \qquad (1)$$

Here, $\sigma_y$ is the yield strength of the material; $\sigma_0$ and $K_f$ are the material constants and $d$ is the average grain size. This equation explains that material strength increases with the decreasing grain size. Interestingly, for ultra-fine grained materials, commonly known as nanocrystalline materials (polycrystal with a grain size less than 100 nm), this effect is inversed resulting in invalidation of common notion "Smaller is stronger"[22]. This phenomenon is called inverse Hall-Petch effect. Conrad et al.[23] reported that the grain size softening or inverse Hall-Petch effect is observed in the nanocrystalline material of grain size less than a critical value. The authors reported a range of 10-50 nm as the critical value for ductile materials. These two phenomena are also observed in 3D ductile materials for glissile nature of dislocations in them. However, numerous studies[20,24] have predicted similar phenomena for single atom thick 2D materials like graphene. The underlying physics of these two cases are distinct, and hence for 2D materials, this is called pseudo Hall-Petch behaviour.

For single crystal materials, the fracture behaviour of a material with pre-existing crack has been analyzed extensively. Due to the stress concentration at the crack tip, the failure always initiates from the crack[25,26]. Multiple grains in a material add uncertainty to the nature of crack propagation. Often times, nanocrystalline materials with pre-existing crack may fail far away from the crack, typically at the grain boundary (GB) junction[27–29]. This resistance to crack propagation or failure from the initial crack tip is coined as flaw tolerance of the material. Zhang et al.[29] showed that nanocrystalline graphene ribbons can be flaw tolerant when the ribbon width is smaller than a critical value of 17.16 nm. Length scale, besides the presence of grain and crack, also affects properties of any new material studied through simulations[30,31]. Properties of nano-materials considerably differ from their bulk manifestation. One such example is fracture toughness of materials. Yin et al.[32] reported that Griffith criterion, based on continuum theory, is not valid for crack sizes less than 10 nm in graphene. In this regard, molecular dynamics (MD) simulations can be used for a comparative study to quantify the amount of deviation in fracture toughness from that of Griffith's criterion.

Therefore, the gap in the understanding of the impact of grain size on fracture strength, fracture toughness, and crack propagation of nanocrystalline silicene (nc-silicene) necessitates a comprehensive study. In this article, we investigate the effect of grain size and crack length on the mechanical properties of nc-silicene under uniaxial tensile loading using MD simulations. Furthermore, we compare the fracture toughness of nanocrystalline materials with Griffith's theory to illustrate its limitation in describing nano-materials' fracture properties. Finally, our findings show that the failure of nc-silicene can be flaw intolerant (crack sensitive) or flaw tolerant (crack insensitive) depending on the crack and grain size as well as the interaction between the cracks and grains.

## 2. Simulation Results and Analysis

To study the mechanical properties of nc-silicene, a 30 nm×30 nm nc-silicene sheet with randomly oriented grains is constructed by Voronoi tessellation method[33] (see Fig. 1). The grain centres and crystallographic orientations are seeded randomly for each grain structures inside the simulation cell. Different initial seeds of the grain structures resulted in different sizes of the grains in the material (Details of grain size calculation are in Supplementary Information). The size effects of the nc-silicene grains on mechanical properties are studied using average grain sizes of 2.5 nm, 5 nm, 10 nm, 15 nm, 17.32 nm, and 21.2 nm under uniaxial tensile loading at a constant temperature of 300 K. Statistical analysis is performed on 10 different samples for each cases to quantify the uncertainties stemming from the randomness of the grain orientations (see Supplementary Information). The uncertainties are incorporated as error bars in the result of this article.

In order to validate our approach, Young's moduli of single crystal and nc-silicene are calculated using SW potential[34]. Our calculated Young's modulus of a single crystal silicene, 85.3 GPa, is in excellent agreement with the reported literature value of 82.2 GPa[35]. We also carried out biaxial tensile loading simulations at 300 K on nc-silicene sheet of 48 nm×48 nm size with a grain size of 8 nm to validate our method. The Young's modulus calculated from this simulation is 143.7 GPa agrees well with the reported value of 136.3 GPa[36] by Liu et al. Further validation of SW potential regarding grain boundary energetics, vacancy formation energy, etc. is provided in Supplementary Information. These calculations establish the credibility of the SW force field in describing mechanical properties of the nc-silicene. Stability of grain boundary is an important aspect in studies involving nanocrystalline structures. We performed grain boundary energy calculations at room temperature in order to evaluate the relative stability of various grain sizes considered in this study (refer to supplementary Fig. S3). The results suggest an increasing trend in the stability of nc-silicene structures with the increase in the grain sizes and 17.32 nm grain size as the most stable grain size. After 17.32 nm, the stability decreases as the grain size increases.

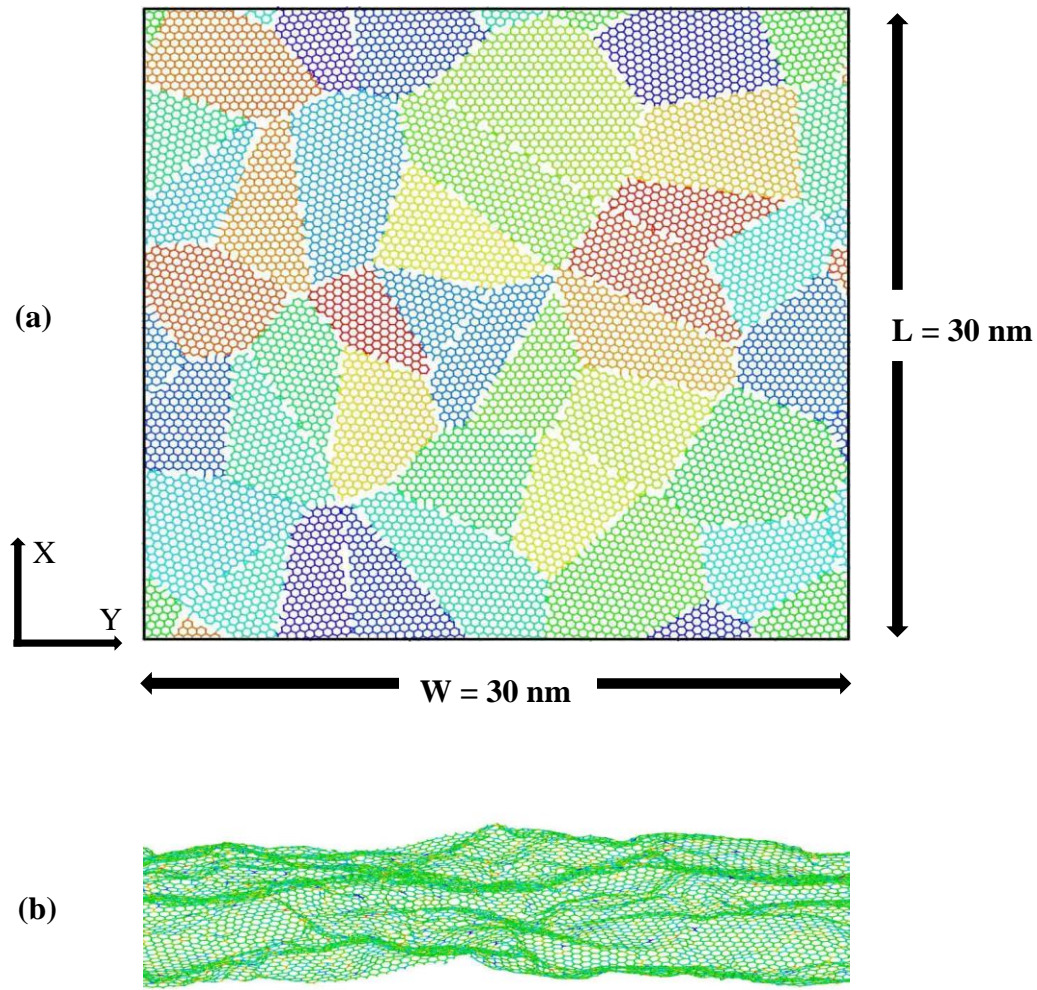

**Figure 1: Atomic configuration of nanocrystalline silicene of average grain size of 10 nm at room temperature (300 K).** (a) Initial geometry of nanocrystalline silicene sheet of 30 nm×30 nm size with randomly oriented grains. The grain boundaries are generated by Voronoi tessellation method. Each grain structure is assigned different colour. (b) After relaxation, the structure shows out of plane deformation due to various defects along the grain boundaries.

Stress vs strain behaviour of single crystal and nc-silicene obtained from tensile test simulations are presented in Fig. 2. The figure reveals that the behaviour of the stress vs strain curve resembles that of 2D polycrystalline graphene sheet obtained from MD simulations[24]. For nc-silicene, the stress initially increases non-linearly up to a certain strain and thereafter increases linearly with additional strain until the failure. This non-linearity stems from the entropic elastic behaviour of nc-silicene sheet. Initially, there are out-of-plane deformations (wrinkles) in the nc-silicene sheet due to grain boundaries. These out-of-plane deformations are known to help minimize the 2D structures[37,38]. Once the applied stress flattens out the sheet, the bonds of silicene get stretched linearly with stress. We observe brittle failure for both single and nc-silicene. However, there is no significant non-linearity in the elastic behaviour of single crystal silicene.

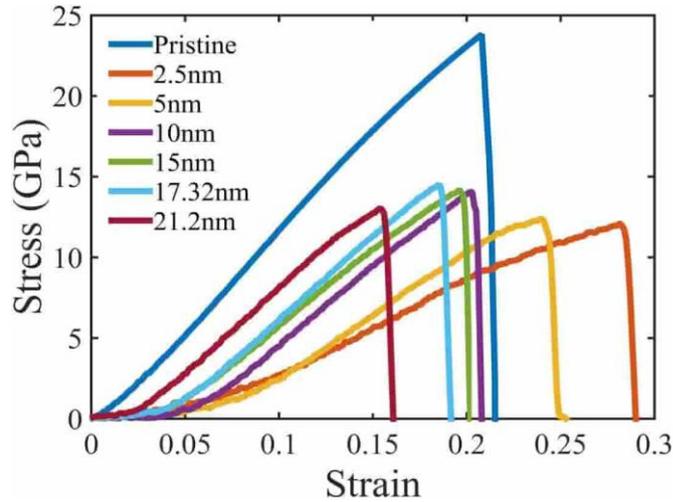

**Figure 2: Stress-strain relation of pristine silicene and nc-silicene of different grain sizes under uniaxial tension at 300K temperature at a loading rate of $10^{-9}$ s$^{-1}$**. The reduced gradient of the curve at the early stage is due to the presence of out of plane deformation in the initial structure after relaxation.

One can see that in Fig. 3(a), the fracture stress of nc-silicene has a well-defined relation with the grain size at a particular temperature. Here, the fracture stress corresponds to the peak stress of the stress-strain curve from Fig. 2 as nc-silicene is found to exhibit brittle failure. The brittle breaking of nc-silicene resembles the ductile fracture of polycrystalline materials following the Hall-Petch and inverse Hall-Petch relations. However, the original Hall-Petch behaviour initiates from the motion of dislocations and grain boundary sliding which are not present in the case of nc-silicene. The reason can be attributed to the brittle failure of the nc-silicene. In light of these contrasting facts, this behaviour in nc-silicene can be termed as pseudo Hall-Petch and inverse pseudo Hall-Petch behaviour. Fig. 3 (a) shows that enlargement of the grain size at nano-scale improves the strength up to a critical grain size and further enlargement causes decrease in the strength of nc-silicene sheet. The critical grain size is observed as 17.32 nm. Grain boundary energy analysis also supports the conclusion of 17.32 as the critical grain size (refer to Supplementary Information). However, this value depends largely on the orientation of grains, the direction of applied stress with respect to the grain, etc. The Pseudo Hall-Petch behaviour can be explained by the stress field created by pentagon-heptagon defects and other dislocations created along the grain boundaries. These dislocations create a repulsive stress field along the grains or a stress barrier. The applied stress must overcome that barrier in order to cause a fracture. As the grain size is reduced the number of dislocations increases (see Fig. 3(d)) to strengthen the sheet.

However, this argument cannot explain the inverse pseudo Hall-Petch region. To elucidate inverse pseudo Hall-Petch behaviour, we are required to examine the weakest-link model for brittle fracture of material. Weakest-link model suggests that the failure strength of a brittle material follows a power-law relation with the number of weak links in the material[39]. The inherent assumption of the weakest-link hypothesis is that a brittle material is comprised of smaller non-interacting sub-structures and the original material fails as soon as the sub-structure or weakest link fails. According to this model, the defects along the grain boundaries, especially in the triple junctions, act as a source of crack or void growth or weak links. . Fig. 3(c) suggests that as the grain size of the nanocrystalline silicene sheet is decreased, the density of triple junctions increases. By fitting the points in Fig.3(c) we observe a power-law relation between the densities of triple junctions with grain size. As the grain size is reduced below a critical value, i. e., into the inverse pseudo Hall-Petch regime, high density of triple junction causes the material to become weaker, thus acting as the weakest-links. In this context one might argue that triple junctions are also present in the coarse grained structures where physical outcomes are opposite. However, for larger grain sizes, impact of pentagon and heptagon defects supersedes the effect of triple junction. We have obtained the values of the constants $\sigma_0$ and $K_f$ of equation (1) by calculating the slope of the straight line in pseudo inverse Hall-Petch region of Fig. 3(a) and its intercept on the axis of fracture stress respectively. In this way, we have derived the following mathematical equation similar to the Hall-Petch relation as in equation (1) to quantify the fracture stress as a function of grain size:

$$\sigma_f = 15.91 - \frac{6.476}{\sqrt{d}}, \tag{2}$$

where, $d$ is in nm and $\sigma_f$ is in GPa. Variation of Young's modulus with grain size (see Fig. 3(b)) exhibits that Young's modulus has a positive relationship with the grain size. As the grain size increases, out-of-plane deformations or wrinkles in the sheet are reduced making the sheet stiffer like the brittle materials. Smaller grain size (large number of grains) actually makes the sheet more compliant to the applied load by creating wrinkles in the sheet.

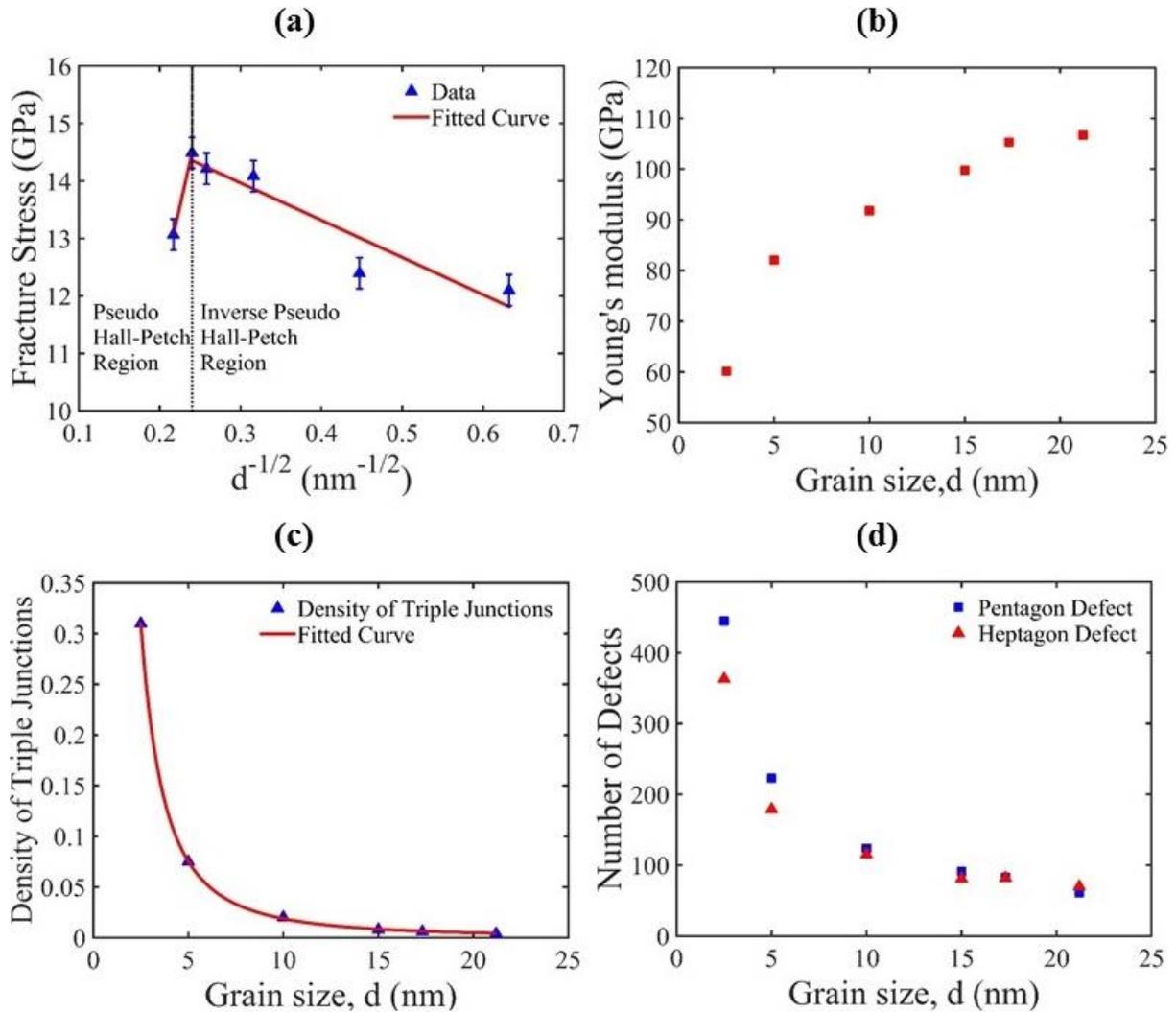

**Figure 3: Variation of (a) fracture stress, (b) Young's modulus, (c) density of triple junctions, and (d) number of defects along the grain boundary of nc-silicene with average grain sizes (expressed as *d*).** In Fig. 3(a), fracture stress follows inverse pseudo Hall-Petch relation up to grain size 17.32 nm. Then it falls into pseudo Hall-Petch region. The "Fitted Curve" in Fig. 3(a) is obtained by fitting the "Data" corresponding to fracture stress.

In this study, the pre-cracked sheets of nc-silicene for different crack lengths are also simulated under uniaxial tensile loading in the directions perpendicular to the direction of crack length. The cracks in the nanocrystalline sheets are formed at the centre by removing multiple lines of atoms to maintain the shape of Griffith's crack. The variation of fracture stresses and strains with the increase of crack lengths are shown in Fig. 4. These graphs are plotted with an error bar obtained from 10 different samples for each case. This approach allows us to quantify the fluctuations in the fracture stress and strain values due to the randomness of grain boundaries.

It is evident from Fig. 4 that both fracture stress and strain decrease with the increase of crack length irrespective of the average grain size. The local stress concentration at the crack tip is more prominent for larger crack lengths. Thus, the atoms near the crack tip undergo an irreversible deformation which promotes faster bond breaking at the crack tip. Due to brittle nature of nc-silicene, once a bond is broken, the failure becomes imminent. It is also observed that the fracture stress of pre-cracked sheets reduces with the decrease in grain sizes. Despite the existence of some discrepancies owing to the random grain boundary (GB), their orientation and the nature of defects (Heptagon-pentagon defects) in the GB, the patterns of stress and strain with the variation of crack length are quite analogous in each case.

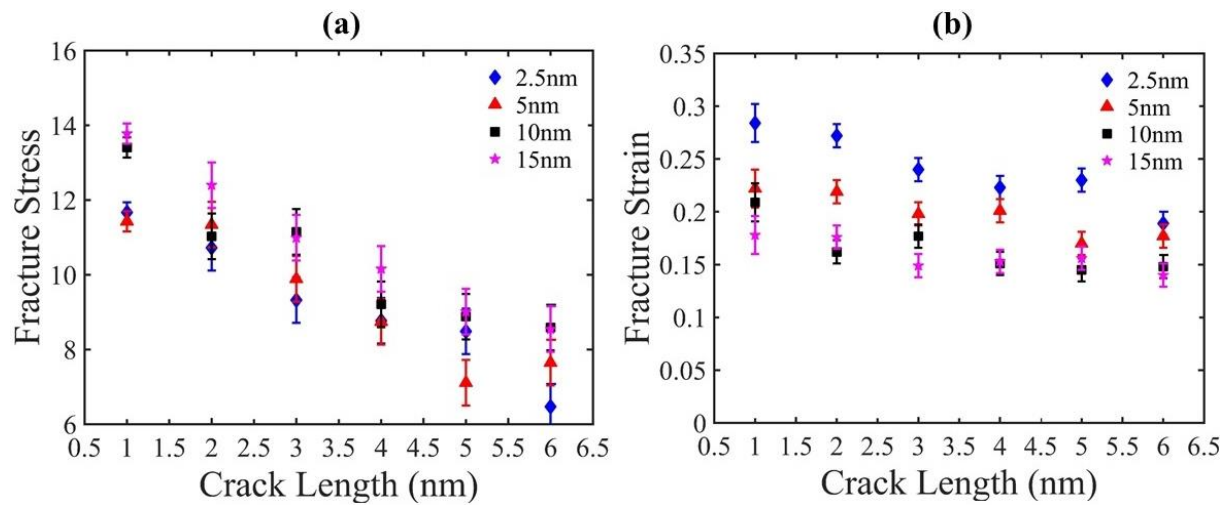

**Figure 4: Variation of (a) fracture stress and (b) fracture strain for different crack lengths of 2*a* = 1-6 nm under uniaxial tension at 300K for various grain sizes.** Both fracture stress and strain show a decreasing trend with crack length.

We also observe an anomalous decrement in the fracture stress and strains for a few cases with larger pre-cracks. This can be attributed to the position of cracks with respect to the GB and the interaction between them. When the stress concentration region produced by the grain boundaries intersects the crack in a direction almost perpendicular to the direction of applied loading, the failure initiates faster. Therefore, the actual fracture stress and strain, in some cases, become lower than the expected value.

Our simulation results for pre-cracked nc-silicene sheet clearly indicate a local stress concentration at the crack tip. According to the classical theory of stress concentration, the failure is always expected to be initiated from the crack tip. However, the nc-silicene behaves quite differently. Often times fracture occurs at the crack, or at the triple junction of the grain boundaries contradicting the prediction of the classical theory of fracture mechanics. Our findings show that the crack sensitivity and insensitivity depends on the relative size and shape of crack and grains as well as their interactions. Thus we observe two types of fracture mechanisms in the present study— crack sensitive fracture and crack insensitive fracture.

To verify the suggested mechanisms of fracture of nc-silicene, the normalized tensile strength is calculated which is defined as[29]

$$\bar{\sigma} = \frac{\sigma_m(a)}{\sigma_t(a)}. \tag{3}$$

Here, $\sigma_m(a)$ is the strength of nc-silicene without any crack and $\sigma_t(a)$ is the limiting strength of the cracked sheet of nc-silicene, where

$$\sigma_t(a) = S(1-\phi). \tag{4}$$

Here, $S$ is the strength of the pre-cracked strip and $\Phi = W/2a$; $W$ is the width of the strip and $2a$ is the length of the center crack.

In Fig. 5, the results of normalized strength are shown as a function of different crack lengths for various average grain sizes. Value of the normalized strength closer to 1 indicates the fracture mechanism is crack insensitive. Fig. 5 depicts that the failure is more likely to be crack insensitive for smaller crack lengths (2 nm and 1nm). Furthermore, in our simulations, we observed for 1 nm and 2 nm crack lengths that failure is crack insensitive for all grain sizes with an exception of sample with 10 nm grain size and 1 nm crack length. Therefore, it is difficult to suggest a critical crack size for crack insensitivity due to the randomness and orientation of grain boundaries (GBs). This randomness in orientation also governs the interaction between GBs and cracks which dictate the failure mechanism. The trend of the graph in Fig. 5 also shows that with the increase of crack length, the normalized strength decreases as nc-silicene exhibits an increased probability of failure from the crack tip, suggesting more crack sensitive fractures.

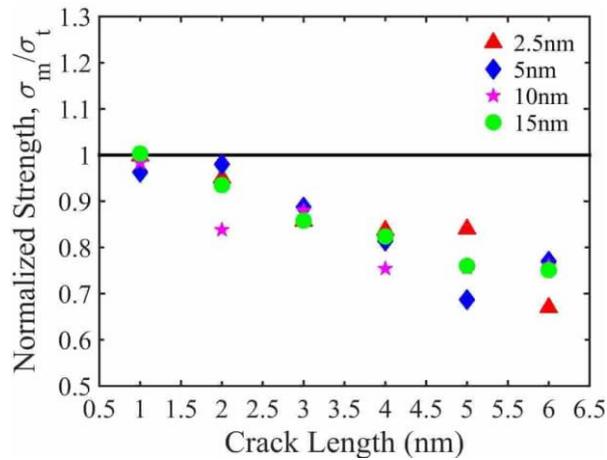

**Figure 5: The normalized strength of nanocrystalline silicene at different crack lengths for an average grain size of $d$ = 2.5 nm, 5nm, 10 nm, and 15 nm.** Here, $\sigma_m$ denotes the strength of nc-silicene without any crack. $\sigma_t$ is the limiting strength of the cracked sheet of nc-silicene. If the normalized strength is closer to 1, failure for the sample is crack insensitive.

To elucidate the mechanism of crack insensitive failure, stress distribution of nc-silicene with average grain size of 15 nm with a crack length of 1 nm is shown in Fig. 6. The figure also presents the steps of crack insensitive deformation process of nc-silicene. To maintain the crack length, the crack in this case is made blunt and almost circular. In Fig. 6(b), the stress concentration region is visible near the crack tip as well as at grain boundaries. However, the stress concentration region near the crack tip remains within the single crystal structure and is also weaker due to the roundness of the crack tip. This single crystal structure is known to be stronger and to break the Si-Si bond, the maximum stress required is reported to be more than 25 GPa[40]. As a result, despite the presence of a stress concentration region at the crack tip, the maximum atomic stress at the crack tip is much lower than the fracture strength required to break a bond near blunt circular crack. Blunt crack tip also makes the crack propagation energetically less favourable. This explains why the failure is not initiated from the crack. On the other hand, the GBs are much weaker than the pristine crystalline structure, and triple junctions are also energetically favourable for rupture with localized atomic stress values exceeding 25 GPa. Therefore, failure is more likely to occur from triple junction instead of the crack tip contradicting the classical theory of crack propagation. This physical reasoning justifies the results shown in Fig. 6.

From atomistic simulations, it is found that the crack insensitive type of failure is dependent on two conditions related to the relative size and shape of cracks and also on the interaction between the GBs and crack. All of these conditions must have to be fulfilled for the occurrence of crack insensitive fracture. The conditions for crack insensitive failure are: (i) the crack size must be smaller than the average grain size, (ii) the crack must be blunt and nearly round shaped, and (iii) the stress concentration region produced by the crack must not interact with the stress concentration region of grain boundaries.

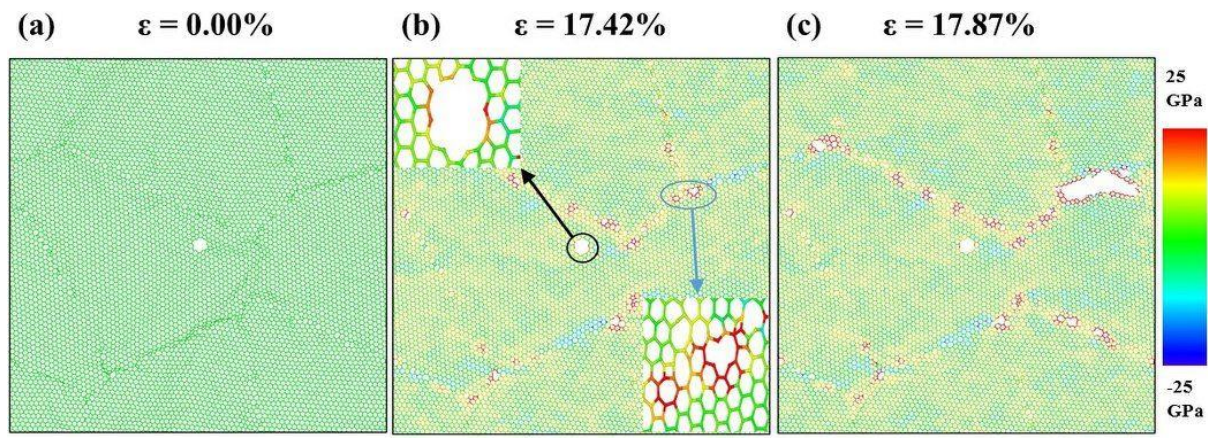

**Figure 6: Atomistic configurations of crack insensitive fracture.** (a) The initial atomic configuration of nc-silicene of 15 nm average grain size with a crack length of 1 nm. (b) Distribution of atomic stress just before the fracture. Here, the black circle indicates a triple junction which shows higher stress concentration and is weaker than the crack. The blue circle indicates that the stress concentration region at the crack tip is inside a pristine crystal ("zoomed-in" views of both the indicated regions are given). (c) Failure from the triple junction. (Another demonstration of crack insensitive fracture in case of nc-silicene of 2.5 nm grain size with 1 nm crack length is shown in the supplementary video SV1)

The atomic stress distribution related to the crack sensitive deformation process of nc-silicene with an average grain size of 2.5 nm and a crack length of 4 nm is shown in Fig. 7. The figure indicates a sharper crack tip, significant stress concentration near the crack tip, and the stress concentration region does not remain localized within a single grain. Therefore, the failure is initiated at the crack tip. Since the crack length is more than the average grain size, the stress concentration region is not localized within the same pristine crystalline structure. This results in a crack sensitive failure. Fig 8 shows another case of crack sensitive failure of nc-silicene sheet with an average grain size and a crack length of 5 nm. Here, the stress concentration region from a grain boundary intersects the stress concentration region of the sharp crack tip, and the intersection occurs in a direction almost perpendicular to the loading direction as shown in Fig. 8 (b). According to previous studies[41–43], the crack propagates along the crack edges and in the direction perpendicular to the applied load because it is energetically favorable. Since the stress concentration region of the GB interacts with that of the crack almost along the crack edges, it favors the crack propagation in that direction. Moreover, the interaction between stress concentration region of the GB and crack tip increases the intensity of local stress at the crack tip initiating a faster bond breaking than the usual crack sensitive fracture. Further bond breaking near the crack tip has made the tip sharper which speeds up the crack propagation. Therefore, for these cases, pre- cracked nc-silicene is found to fail at a stress lower than the usual crack sensitive fracture. This phenomenon is also evident in the anomaly in the decrement pattern of strains and stresses with the increase of crack length as shown in the Fig. 4. This irregularity can be attributed to the tendency of GBs to favor the nucleation and propagation of crack tip.

The aforementioned analysis implies that crack sensitive failure in nc-silicene occurs under three conditions. If the system can satisfy any of these conditions, it becomes crack senstive. These three conditions are: (i) the crack length is higher than the average grain size, or (ii) the crack is not round shaped, or (iii) the crack is smaller than average grain size, but the stress concentration region produced by GBs interact with the stress concentration region of the crack, especially in the direction of crack edges as shown in Fig. 8.

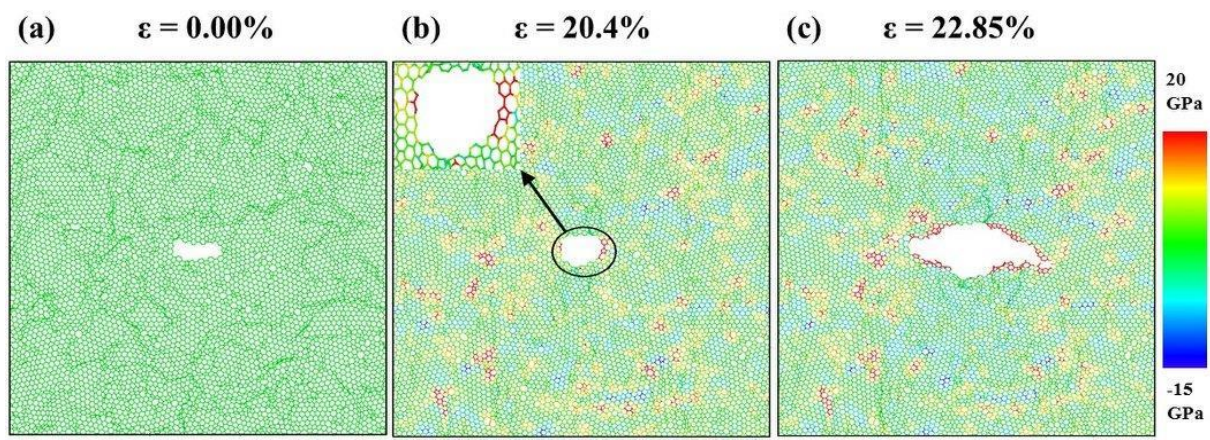

**Figure 7: Atomistic configurations of crack sensitive fracture.** (a) The initial atomic configuration of nanocrystalline silicene of 2.5 nm grain size with 4 nm length of the crack. (b) Distribution of atomic stress before the fracture showing strong stress concentration at crack tip indicated by black circle ("zoomed-in" view is given). (c) Failure from the crack tip.

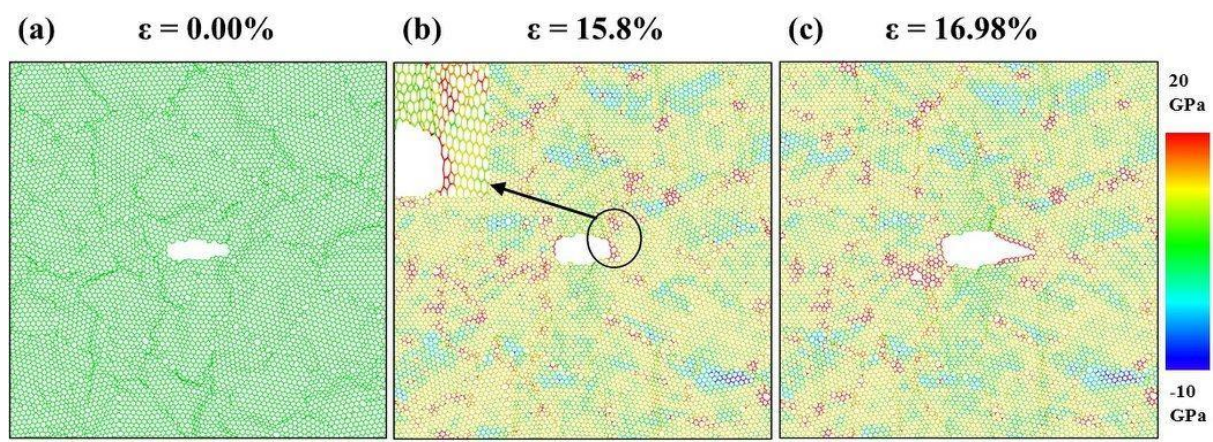

**Figure 8: Crack sensitive fracture due to overlapping of stress concentration between crack tips and GBs.** (a) The initial atomic configuration of nc-silicene of 5 nm grain size with 5 nm crack length. (b) Distribution of atomic stress before the fracture. Here, black circle indicates overlapping of stress concentration region of crack tip and GB ("zoomed-in" view is given). (c) Failure from a crack tip from the side of overlap.

In our analysis, the samples of nc-silicene with crack length more than 2 nm are found to fulfill any of the above mentioned conditions. Thus, they experience the crack sensitive fracture phenomenon. To further explore the crack sensitive and insensitive mode of fracture, we simulated the cases demonstrated in Figs. 6-8 at lower strain rates of $10^8 \, s^{-1}$, $5 \times 10^8 \, s^{-1}$ and time step of 0.0005 ps to check whether their fracture behaviour remains the same or not (the results are included in the Supplementary Information). It is found that the crack sensitive and insensitive fracture exists irrespective of the strain rate and timestep.

We have studied the fracture toughness of nc-silicene from MD simulations and Griffith's theory. The fracture toughness measures the ability of a pre-cracked material to resist its failure. The theoretical formulation of fracture toughness is made by the classical Griffith's equation for a central crack[44]. According to this equation, the critical stress for fracture of a stripe with finite central crack is given by-

$$\sigma_f = \frac{1}{f(\Phi)}\sqrt{\frac{E\Gamma}{\pi a}}. \tag{5}$$

Here, $E$ is the Young's modulus, $\Gamma$ is the surface energy for 3D material; edge energy for 2D material; crack formation energy for polycrystalline materials, and $a$ is half of the crack length, and the function $f(\Omega)$ is a geometrical factor given by-

$$f(\Phi) = [1 - 0.025\Phi^2 + 0.06\Phi^4] \times \text{Sec}\left(\frac{\pi\Phi}{2}\right)^{1/2}, \tag{6}$$

where $\Phi = 2a/W$; $W$ is the width of the strip with a central crack of length $2a$. The edge energy in equation (5) is the difference of the energy released by the nc-silicene samples with and without cracks. Edge energy is a constant for a nanocrystalline material of same average grain size, GB and crack orientations irrespective of the crack size[45]. The calculated edge energies obtained in our analysis for different crack lengths are in close proximity. Therefore, an average value of $\Gamma$ for all crack lengths is considered. By calculating the edge energy, the fracture toughness is determined using the following equation derived from equation (5)-

$$\sigma_f \sqrt{a} = \frac{1}{f(\Omega)}\sqrt{\frac{E\Gamma}{\pi}}. \tag{7}$$

Furthermore, the fracture toughness of nc-silicene obtained from MD simulations is calculated by $\sigma_f\sqrt{a}$, where $\sigma_f$ is the fracture stress of the pre-cracked sheet. The fracture toughness calculated using Griffith's equations and molecular dynamics simulations are presented in Fig. 9 for various cracks lengths and average grain sizes. The fracture toughness predicted by MD simulations are plotted with error bars to represent the uncertainties associated with the randomness of GB distribution. The figures show that the MD simulation results deviate significantly from the fracture toughness obtained from Griffith's equations. This deviation is particularly stemmed from the limitations of theory of Griffith —which is based on continuum assumptions—in describing nano-crystalline materials. Moreover, the assumption of a perfectly elliptical crack in the Griffith's theory cannot be maintained in the nc-silicene sheets due to its hexagonal structure. It is also worth mentioning that Griffith's theory is governed by the local strength of the atoms at the crack tip, while the MD simulation results focus on global energy balance according to virial stress theorem[46]. The limitation of the Griffith theory to describe nano-materials has been reported in previous studies. Yin et al. reported that the Griffith's theory failed for graphene nano-sheets with a crack size smaller than 10 nm[32]. The validity of Griffith's theory, as in the current study, is also limited for nc-silicene with nano-sized crack. The critical crack size for the limitation of Griffith's theory in describing nc-silicene will be explored in a future study.

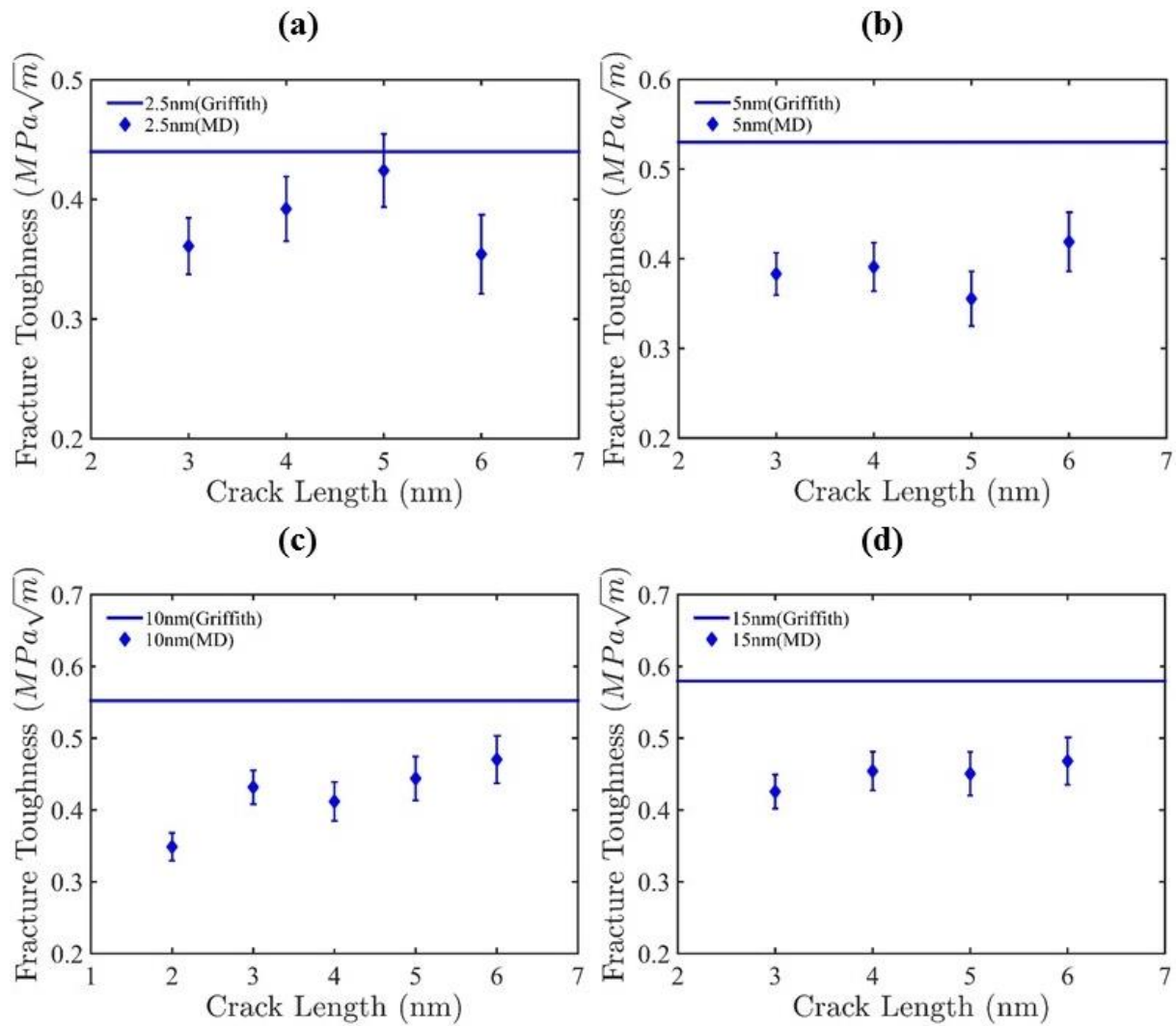

**Figure 9: Prediction of fracture toughness by Griffith's criterion and molecular dynamics at different crack lengths for grain size of (a) 2.5 nm, (b) 5 nm, (c) 10 nm, and (d) 15 nm.** Fracture toughness obtained by Griffith's theory is constant for all the crack lengths because the edge energy $\Gamma$ is averaged for all crack lengths and assumed as constant. So, theoretical fracture toughness is an average prediction.

The pseudo inverse Hall-Petch effect depicts that with the increase of average grain size, the fracture stress is increased as found in nc-silicene (shown in Fig. 3). This is intriguing that the fracture toughness as a function of grain size shows the similar trend. Both theoretical and MD results of fracture toughness are found to increase with the increased average grain size. This observation explains that the cracks in a nanocrystal with a larger grain size resist failure more than that of the nanocrystal with a smaller grain size.

Furthermore, in a crack insensitive fracture, the failure is not initiated from the crack tip. Rather, it fails from a triple junction and behaves more like nc-silicene sheet without a crack. According to the definition of fracture toughness, it is the ability of a material to resist crack propagation. Since the crack insensitive fracture physically demonstrates a fracture mechanism like nc-silicene sheet without crack, the ultimate stress of crack insensitive fracture cannot depict fracture toughness. Thus, these are not incorporated in Fig. 9.

It is also important to note that the results of fracture toughness reflect greatly on the anomalous decrement of fracture stress observed in Fig. 4. Due to the position of the crack tip with respect to GB, there is an interaction of stress concentration region between them. This leads to an increase in the local stress which causes the bond breaking faster than the usual. Therefore, the fracture stress is reduced than the usual value and as a result, there is also a decrease in the fracture toughness as obtained from our simulations. This phenomenon is evident in the cases of 2.5 nm grain size with 5 nm crack length, 5 nm grain size with 5 nm crack length, and 10 nm grain size with 4 nm crack length, and 15 nm grain size with 5 nm crack length.

### 3. Discussions

In this study, we found that the mechanical properties of nc-silicene are contingent on average grain and crack sizes. We also elucidate the failure mechanism of nc-silicene and demonstrate the deviation of classical theories at nano-scale. Our simulation results reveal that nanocrystalline silicene fails at lower values of ultimate stress than the single crystal silicene due to the presence of grain edges. We observe brittle failures in both single and nc-silicene. The fracture stress of nc-silicene increases with the increased grain size contradicting classical Hall-Petch effect up to a critical grain size. This phenomenon is termed as inverse pseudo Hall-Petch effect. We found the critical grain size as 17.32 nm above which the fracture stress follows the pseudo Hall-Petch effect. Typically, fracture stress and strain show decreasing trends with the increase of crack length. However, in some cases, due to the interaction between stress concentration region of crack and GBs, the failure may occur at a lower stress than the usual values. One of our key findings is that nc-silicene exhibits two types of fracture mechanisms, namely, crack insensitive and crack sensitive fracture. In crack insensitive fracture, fracture proceeded from the triple junction of GBs even in the presence of a crack. Crack insensitive fracture occurs when the crack length is less than the average grain size and the stress concentration region of GB does not interact with the stress concentration region of crack. On the other hand, the crack sensitive fracture occurs at the crack tip as predicted by the classical theory. We observe that the fracture is crack sensitive when the crack length is more than the average grain size. Furthermore, nc-silicene also experiences crack sensitive failure even with a crack length that is smaller than the average grain size. It occurs due to the interaction of GB stress concentration region with that of crack, especially in the direction along the crack edges. Finally, we observe that the fracture toughness of nc-silicene calculated from MD simulation deviates from fracture toughness obtained from the Griffith's theory. This deviation limits the applicability of Griffith's theory at nano-scale. Griffith's theory also over-predicts the fracture toughness than that of MD simulations. This prediction is contrary to the case of single crystal silicene because of the higher energy release from the defects at the crack edges in nc-silicene than in single crystal silicene. We believe that this study paves a way to comprehensive and deeper understanding of the mechanical properties and failure mechanism of nc-silicene which is regarded as one of the future materials for manufacturing nano-electronic devices.

## 4. Methods

MD simulation method has been established as an effective way to explore deformation mechanism and fracture of polycrystalline materials at atomic scale[47–49]. In this study, we used an optimized Stillinger-Weber (SW) potential to describe the atomic interactions of silicon atoms. Previously, this potential has been used for describing thermal[34] and mechanical properties of single crystal silicene[40]. The potential is validated against the structural properties calculated by Density Functional Theory (DFT)[50]. Moreover, our structural relaxation simulations reproduce the buckled structure of silicene. The buckling height obtained is 0.44 Å which is reasonably in good agreement with DFT calculations[50]. The SW potential consists of two terms; a two-body term describing the bond stretching interactions and a three-body term describing the bond breaking interactions. These interactions are expressed as follows

$$\Phi = \sum_{i<j} V_2 + \sum_{i<j<k} V_3, \tag{8}$$

$$V_2 = \varepsilon A (B \sigma_a^p r_{ij}^{-p} - \sigma_a^q r_{ij}^{-q}) e^{[\sigma_a (r_{ij} - a_1 \sigma_a)^{-1}]}, \tag{9}$$

$$V_3 = \varepsilon \lambda e^{[\gamma \sigma_a (r_{ij} - a_1 \sigma_a)^{-1} + \gamma \sigma_a (r_{jk} - a_1 \sigma_a)^{-1}]} (\cos\theta_{ijk} - \cos\theta_0)^2, \tag{10}$$

where, $V_2$ and $V_3$ are two-body and three-body terms respectively; $r_{ij}$ is the distance between atoms $i$ and $j$; $\sigma_a$ is the finite distance at which the inter particle potential is zero; $\theta_{ijk}$ is the angle between bond $ij$ and $jk$; $\theta_0$ is the equilibrium angle between two bonds; other parameters like $a_1$, $A$ and $B$ are the coefficients required to fit when developing the potential. The values of these parameters are shown in Table 1:

**Table 1: Optimized SW potential parameters for silicene.** Here, the unit of ε is in electron volts and $\sigma_a$ is in Å. The other parameters are dimensionless and their definitions are given in the paper[34].

| ε (eV) | $\sigma_a$ (Å) | $a_1$ | λ | Υ | $Cos\theta_0$ | A | B | p | q | tol |
|---|---|---|---|---|---|---|---|---|---|---|
| 2.1683 | 1.99751 | 1.8 | 22.275515 | 1.2 | -0.44561011 | 5.834064 | 0.602225 | 4 | 0 | 0 |

We used LAMMPS software package[51] for all the simulations. Periodic boundary conditions are employed for in-plane directions, and the out-of-plane direction is considered as a free surface. The geometries were relaxed by using conjugant gradient minimization scheme. Then the system was equilibrated using NVE simulations, followed by NPT relaxation at 300 K temperature and atmospheric pressure. Next, uniaxial stress was applied at a constant strain rate $10^9$ s$^{-1}$. The equation of atomic motion was integrated with time step 1 fs. We note that the strain rate applied in our simulations is considerably higher than the practical cases. The higher strain facilitates us to explore the material failure mechanisms using a moderate computational resource. We observe an out of plane deformation of nc-silicene sheet after the full relaxation as shown in Fig. 1. Under deformation, the atomic stresses are calculated using the following equation[46]

$$\sigma_{ij}^{\alpha} = \frac{1}{\Omega^{\alpha}} \left( \frac{1}{2} m^{\alpha} v_i^{\alpha} v_j^{\alpha} + \sum_{\beta=1,n} r_{\alpha\beta}^{j} f_{\alpha\beta}^{i} \right), \quad (11)$$

where i and j denote indices in the Cartesian coordinate system; α and β are the atomic indices; $m^{\alpha}$ and $v^{\alpha}$ denote the mass and velocity of atom α; $r_{\alpha\beta}$ is the distance between atoms α and β; $f_{\alpha\beta}$ is the force between atoms α and β; $\Omega^{\alpha}$ is the atomic volume of atom α. Zhou[52] definitively proved that inclusion of the kinetic term (first term in the right side of equation 11) erroneously calculates the stress and hence in this study kinetic contribution of the atoms to stress is neglected. Then the reported stress during deformation is obtained by averaging the atomic stresses of the system.

**Author Contribution:** T.R. and S.S. initiated the idea under the supervision of M.M. and M.M.I.. S.S. generated the structure of nanocrystalline silicene. T.R. and S.S. performed the MD simulations. T.R.

and S.S analyzed the results and prepared the manuscript. S.M. contributed to interpreting the results. All authors contributed to this work.

**Additional Information**

**Competing Financial Interest:** The authors declare no competing financial interests.